\documentclass[epj]{svjour}
\usepackage{amsmath}
\usepackage{latexsym}
\usepackage{graphics}
\newcommand{\aver}[1]{\left< #1 \right>}
\begin{document}
\title{Quantum transport in a driven disordered potential:
onset of directed current and
noise-induced current reversal}

\author{D.~V.~Makarov \and L.~E.~Kon'kov% etc
}                     % Do not remove
\institute{Laboratory of Nonlinear Dynamical Systems,\\
V.I.Il'ichev Pacific Oceanological Institute
of the Far-Eastern Branch of the Russian
Academy of Sciences, \\ 43 Baltiyskaya St.,
690041, Vladivostok, Russia}
\date{Received: date / Revised version: date}
\abstract{
We study motion of a quantum wavepacket in a one-dimensional potential with correlated disorder.
Presence of long-range potential correlations allows for existence of both localized and extended states.
Weak time-dependent perturbation in the form of a fluctuating plane wave
is superimposed onto the potential. 
This model can be realized in experiments with optically trapped cold atoms.
Time-dependent perturbation causes transitions between localized and extended states.
Owing to violation of space-time symmetries, there arises atomic current which is codirectional with the wave-like perturbation.
However, it is shown that the perturbation can drag atoms only within some limited
time interval, and then the current changes its direction.
Increasing of the perturbation bandwidth and/or amplitude results in decreasing of the time of current reversal.
We argue that onset of the current reversal is associated with inhomogeneity of diffusion in the momentum
space.
 % end of PACS codes
} %end of abstract
\authorrunning{Makarov, Kon'kov}
\titlerunning{Quantum transport in a driven disordered potential}
\maketitle
\section{Introduction}
\label{intro}

It is well known that cold atoms trapped in optical lattices
can serve as an excellent quantum simulator of solid-state
physics phenomena \cite{Buluta}.
For example, creation of artificial magnetic fields 
in 2D lattices allow for studying
quantum Hall effect with exceptionally strong magnetic fields
\cite{Kolovsky-EPL11,Aidelsburger-PRL11,Aidelsburger-APB,Ketterle-PRL13,JPB13}
which are hardly achievable in real solid-state experiments.
Quantum ratchets with cold atoms \cite{Carlo06,Gong,Flach07,Gong_Poletti_Hanggi,Morales-Molina_Flach-NJP08,Hanggi_Marchesoni,Poletti09,Ponomarev}
can serve 
as simulators of the related photogalvanic phenomena 
in solid-state nanostructures 
\cite{Belinicher,AlekseevKN,Entin_Magarill,Weber-PRB08,Pyataev_Ulyanov,Tarasenko-PRB11,Kiselev_Golub,Ermann_Shepelyansky-EPJB,Nalitov}. 
Besides, the ratchet effect has self-contained meaning
as a tool for controllable
transportation of atoms into some target region,
that is of great importance in nanoscale technologies like
quantum communication \cite{Romero-Isart-PRA07,Calarco-control}.

From the viewpoint of various solid-state applications,
it is reasonable to examine the possibility
of gaining dc current if the spatially periodic potential
is replaced by a random one.
An example of a classical ratchet with a disordered potential is presented in \cite{Liebchen-PRL14}.
Indeed, it is well known that ballistic transport in 1D undriven disordered potentials
is prevented by scattering processes which can give rise to the Anderson localization. 
However, external AC driving can significantly increase the localization
length or lead to delocalization even within 
the tight-binding approximation
\cite{Holthaus-PRL95,Martinez_Molina-PRB,Martinez_Molina-EPJB} 
that doesn't take into account Landau-Zener interband tunneling.
As number of frequency components in the driving increases,
 the resulting transport transforms from subdiffusive to diffusive 
 \cite{Yamada_Ikeda-PLA98,Yamada_Ikeda-PRE99,Yamada_Ikeda-PhB99,Yamada01,Yamada_Ikeda-PRE02,Yamada_Ikeda-PLA04,Derrico,Cao-NJP13}.
Landau-Zener tunneling results in energy growth that also facilitates 
delocalization \cite{Wilkinson88,Wilkinson91}.
%it is well-known that 
%the Anderson localization can be destroyed by time-dependent driving.
So, one may expect that properly constructed external AC driving should give rise
to directed ballistic current, i.~e.
the ratchet effect, provided certain time-space symmetries of the driving are violated.
This is an important advantage of the ratchet effect
as compared to the action of stationary directed forces:
in the latter case eigenfunctions remain exponentially localized that infers the insulating regime.

In the present work, 
AC driving is superimposed onto smooth potential with correlated disorder. 
We consider AC driving being a superposition of two optical lattices
whose amplitudes are small and subjected to broadband modulation that is modeled using harmonic noise.
With proper choice of the phase shift between the modulating signals,
the driving force becomes a running plane wave experiencing time and space fluctuations.
This kind of ratchets is known as travelling potential ratchets and considered in \cite{JETPL,PRE75,JTPL08,EPJB,PLA,Denisov-review}.
They can be used as quantum simulators of electron-phonon interactions in semiconducting materials \cite{Yamada_Ikeda-PLA96,Jiang_etal,Roche-JP07,Greenaway-PRB}.
This issue is of especial importance as a promising way 
of electronic transport control using the stimulated phonon emission that can be realized via a SASER \cite{SASER02,Kent06}.
Interest to our configuration is substantially supported by recent results of 
\cite{Fromhold-PRA13}, where non-trivial dependence of current on the perturbation strength was found for a somewhat similar system.

The paper is organized as follows. In the next section we give detailed description of the model studied.
In particular, we study spectral properties of the undriven system and point out the presence of the mobility edge
in the energy space. Also, we introduce AC driving involving harmonic noise and describe its basic properties.
Section \ref{Numer} contains results of numerical simulation. 
These results are analyzed in section \ref{Momentum} in terms of kinetic approach.
In section \ref{Summary} we summarize the results and outline ways of further research.

\section{Model description}
\label{Model}

\subsection{Time-independent part}
\label{Undriven}

One-dimensional motion of an atomic wavepacket along the x-direction is governed by the Schr\"odinger equation
\begin{equation}
i\hbar\frac{\partial\Psi}{\partial t}=-\frac{\hbar^2}{2M}\frac{\partial^2\Psi}{\partial x^2}
+ [U(x) + \varepsilon V(x,t)]\Psi,
\label{Shrod} 
\end{equation}
$M$ is atomic mass, $\varepsilon$ is a small parameter. Hereafter we use scaling corresponding to $M=1$.
\begin{figure}[!htb]
\resizebox{0.45\textwidth}{!}{%
  \includegraphics{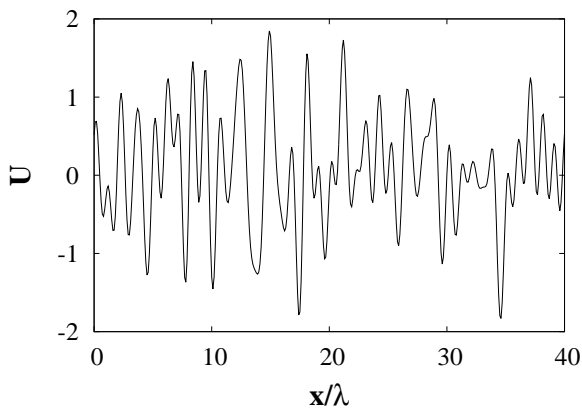}
}
\caption{A typical realization of the potential $U(x)$, $\lambda=2\pi/k_0$.}
\label{fig-speckl}       
\end{figure}
Function $U(x)$ describes the time-independent part ot the optical potential. 
We construct it as superposition of plane waves with random wavevectors and phases:
\begin{equation}
 U = A\tilde U,\quad
 \tilde U=\sum\limits_j \cos(k_0x\cos\theta_j + \chi_j).
\label{U}
 \end{equation}
Here $\theta_j$ and $\chi_j$ are random phases with uniform distribution 
in the interval $[0:2\pi]$, and $k_0=1$. 
Coefficient $A$ is determined by the normalization condition
\begin{equation}
 A = \left(2\aver{\tilde U^2}_x\right)^{-1/2},
\end{equation}
where $\aver{...}_x$ denotes averaging over $x$.
An example of $U(x)$ is presented in Fig.~\ref{fig-speckl}.
According to the figure, $U(x)$ can be regarded as some randomly-distorted lattice potential.
It can serve as a model of an optical potential created by optical speckle pattern.

Autocorrelation function of the potential (\ref{U}) obeys the following formula \cite{MaksSadr08}
\begin{equation}
 \aver{U(x)U(x+d)}\, \propto J_0(d),
\end{equation}
where $J_0$ is the zero-order Bessel function of the first kind. 
Power-like asymtotics of $J_0$ implies the presence of long-range correlations.
Long-range correlations result in the existence of the mobility edge, i.~e. the energy boundary
separating localized and delocalized states \cite{IzrKrokhin}.
A simple intuition suggests that the transition to the delocalization should occur with increasing of energy.
The transition should be reflected in the energy spectrum, in particular, in statistics of level spacings \cite{Sivan_Imry,Altshuler88repulsion,Shklovskii_etal93}
\begin{equation}
 s = \epsilon_{n+1} - \epsilon_{n},
\end{equation}
where $\epsilon_{n}$ and $\epsilon_{n+1}$ are two consecutive unfolded energy values.
Unfolding is the procedure used in order to extract the fluctuating part in the energy level density.
Unfolded energy values are related to the original ones by means of the formula
\begin{equation}
 \epsilon_i = \bar{N}(E_i),
\end{equation}
where 
\begin{equation}
 \bar{N}(E)=\int\limits_{-\infty}^E \bar\rho(E')\,dE'.
\end{equation}
Here $\bar\rho(E)$ is the mean level density being smooth approximation to the actual level density.
In the present work, we use the so-called local unfolding \cite{Gomez-unfolding}, when $\bar\rho(E)$ is 
\begin{equation}
\bar\rho(E_i)=\frac{2\nu}{E_{i+\nu}-E_{i-\nu}},
 \label{local}
\end{equation}
%
%To verify this, one can analyze spectral statistics of the undriven system.
with $\nu=5$.

In the insulating regime corresponding to the disorder-induced localization,
eigenstates belonging to the same energy band
don't substantially overlap in space, therefore, their energies are statistically independent
and obey Poissonian distribution of level spacings
\begin{equation}
P(s) = \frac{1}{\aver{s}}\exp\left(-\frac{s}{\aver{s}}\right),
\label{Poisson} 
\end{equation}
where $\aver{s}$ is the mean level spacing.
Non-zero conductivity implies 
overlapping of eigenstates that gives rise 
to level repulsion, whereby
level spacing statistics is described by the Wigner surmise
\begin{equation}
 P(s) = \frac{\pi}{2}\frac{s}{\aver{s}^2}\exp\left(-\frac{\pi s^2}{4\aver{s}^2}
 \right).
\end{equation}
Finally, consider
the regime of free motion that is not affected by the potential.
In this case energy spectrum inside a sample of length $L$ with perfectly reflecting boundaries
is described by the simple formula
\begin{equation}
 E_n=\frac{2\pi^2\hbar^2}{L^2M}n^2.
 \label{Efree}
\end{equation}
Absence of potential-induced fluctuations implies uniform density of unfolded levels with constant level spacing $s$.
In the model we consider, this regime should be relevant for the range of
high energy values.

%Presence of the mobility edge implies qualitative change of level spacing statistics with increasing energy.
%
\begin{figure}[h!tb]
\resizebox{0.45\textwidth}{!}{%
  \includegraphics{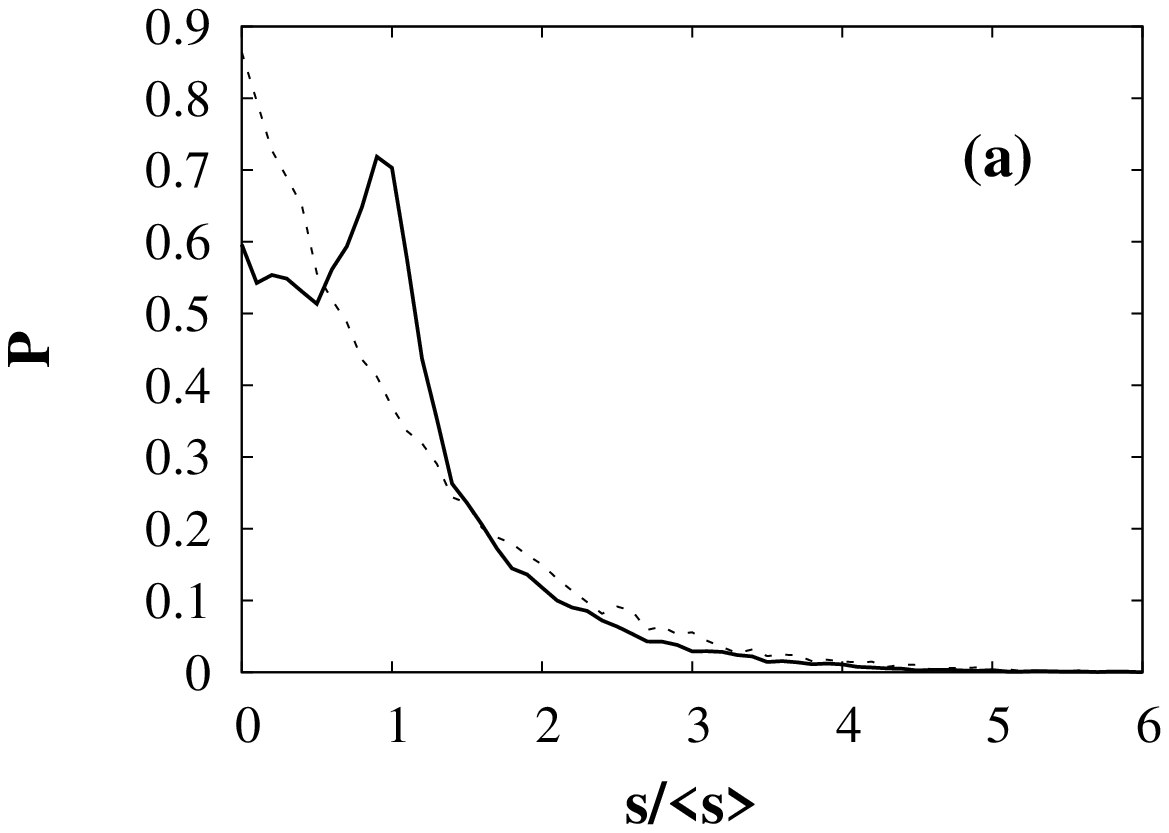}
}\\
\resizebox{0.45\textwidth}{!}{%
  \includegraphics{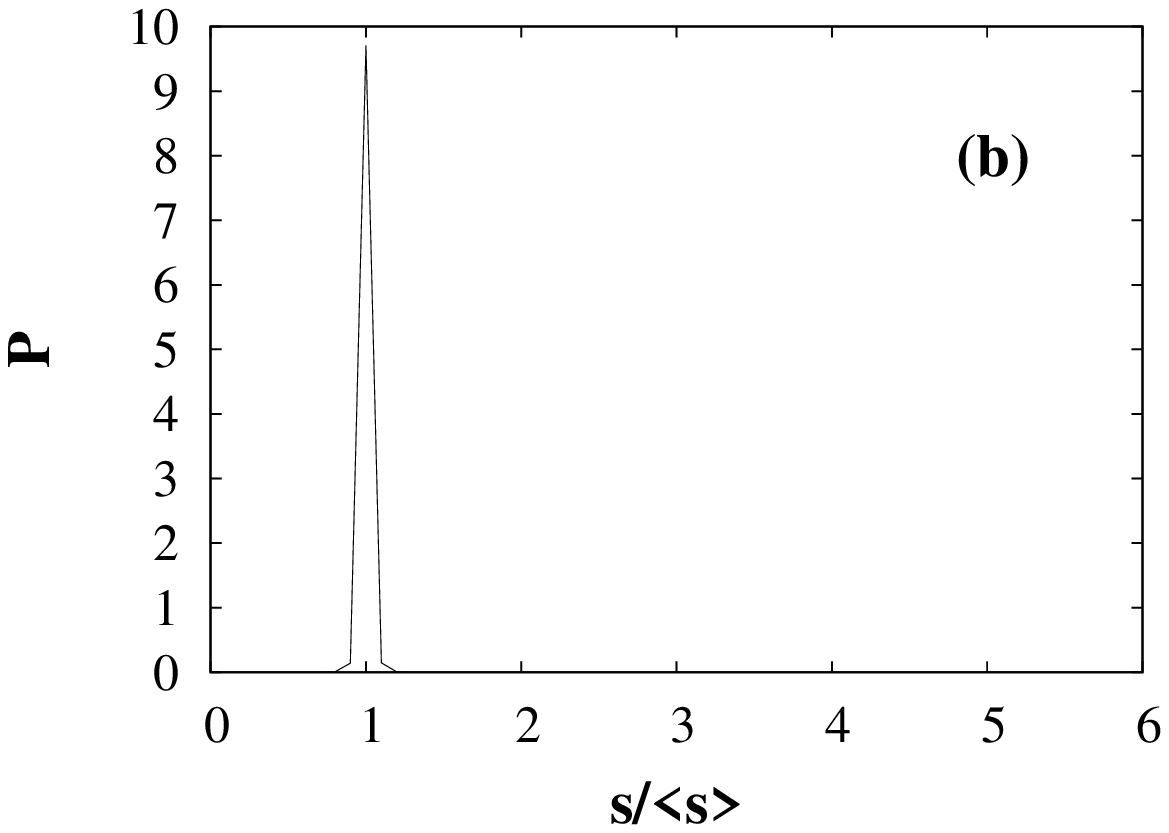}
}
\caption{Level spacing distributions in various energy bands.
(a) $E\le 1$ (dashes) and $1<E\le 4$ (thick solid), (b)  $4<E\le 12$.} 
\label{fig-sp}
\end{figure}

Figure \ref{fig-sp} demonstrates level spacing distributions corresponding to the different
energy bands. 
Level spacing distributions were obtained by solving numerically the stationary Schr\"odinger equation
with $\varepsilon=0$ for a long sample ($L=10000\pi$). 
In the lowest energy region, level distribution is well described by the Poissonian law (\ref{Poisson}),
indicating localization of eigenstates. In the moderate energy region, $1<E\le 4$, level spacing distribution
significantly deviates from the Poissonian form and is non-monotonous, revealing the presence of level repulsion.  
Level spacing statistics in
the high-energy range represents sharp peak at $s=<s>$. It corresponds to the absence of level fluctuations in the ballistic regime.
So, one can deduce the existence of the mobility edge in the range $1<E\le 4$, separating low-energy localized states
and high-energy metallic ones.

\subsection{Time-dependent perturbation}
\label{Drive}

\begin{figure}[h!tb]
\resizebox{0.45\textwidth}{!}{%
  \includegraphics{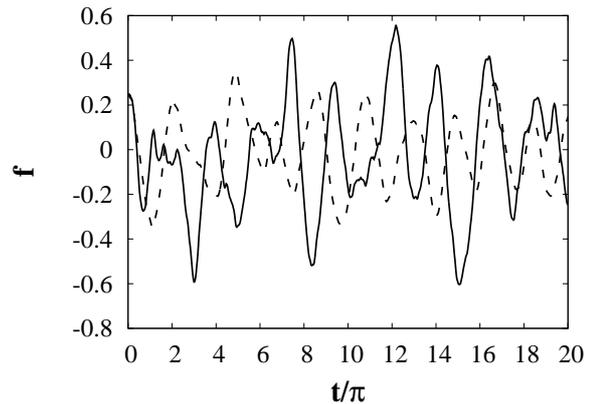}
}
\caption{Realizations of harmonic noise for 
$\Gamma = 0.1$ (dashes) and $\Gamma=0.5$ (solid).} 
\label{fig-hn}
\end{figure}

As is shown in the preceding section, potential (\ref{U}) allows for extended states despite of disorder.
Localized and extended states are separated by the mobility edge.
This gives rise to the possibility for transition between the insulating and conducting regimes 
by means of external time-oscillating driving.
If the driving doesn't satisfy certain space-time symmetry relations,
the resulting current can be directed \cite{FlachYevtushenkoZolotaryuk,Liebchen-NJP12}.
Following \cite{PLA,Order},
we use the perturbation 
 $V(x,t)$ expressed as
\begin{equation}
 V(x,t) = f(t)\sin{x} + f(t+\tau)\cos{x}, 
\label{Vxt}
\end{equation}
where $f(t)$ is a broadband signal. 
Thus, $V(x,t)$ is given by superposition of two lattice potentials subjected
to amplitude modulation. Note that the modulating signal $f(t+\Delta)$ is the replica of the signal
$f(t)$ with the time shift $\tau$.

In experiments, broadband amplitude modulation of optical lattices
can be realized using coherent frequency-modulated signals. 
However, in the present work we consider
a more complicated case, when non-zero bandwidth of modulating signals is associated with
uncontrollable stochastic processes. 
Nevertheless, it is assumed that the modulating signal can be recorded and reproduced.
In particular, we model $f(t)$ as the so-called harmonic noise \cite{HN,Anischenko}.
Harmonic noise is described by coupled stochastic differential equations
\begin{equation}
 \dot f=y,\quad
 \dot y=-\Gamma y-\omega_0^2f + \sqrt{2\beta\Gamma}\xi(t),
 \label{ou2d}
\end{equation}
where $\Gamma$ is a positive constant, and
$\xi(t)$ is Gaussian white noise.
Realizations of harmonic noise can  be calculated by means of mapping
\begin{equation}
 \begin{aligned}
  &f_{n+1}=f_n + s_nh + \frac{1}{2}\alpha_nh^2 + \gamma Z_2(h),\\
  &s_{n+1}=s_n + \alpha_nh + \gamma Z_1(h) - \frac{1}{2}\Gamma\alpha h^2 \\
  &+ \Gamma\gamma Z_2(h) - \frac{1}{2}\Omega^2s_n h^2,
 \end{aligned}
\end{equation}
where $h$ is the time step, $\alpha_n=-\Gamma s_n-\Omega^2f_n$,
$\gamma=\sqrt{2\beta\Gamma}$.
Terms $Z_1$ and $Z_2$ are given by expressions
\begin{equation}
 Z_1(h)=\sqrt{h}Y_1,\quad
Z_2(h)=h^{3/2}\left(\frac{Y_1}{2} + \frac{Y_2}{2\sqrt{3}}
\right),
\end{equation}
where $Y_1$ and $Y_2$ are statistically independent Gaussian noises with unit variance.
Realizations of harmonic noise with different values of $\Gamma$
are exemplified in Fig.~\ref{fig-hn}.

The terms $f(t)$ and $f(t+\tau)$ in (\ref{Vxt})
correspond to one and the same realization of harmonic noise and
differ only by the temporal shift $\tau$.
The first two moments of harmonic noise are given by 
\begin{equation}
 \left<f\right>=0,\quad
 \left<f^2\right>=\frac{\beta}{\omega_0^2}.
\end{equation}
We set $\beta=1$, that is, the perturbation strength is solely determined 
by the parameter $\varepsilon$ entering into (\ref{Shrod}).
In the case of low values of $\Gamma$,
the power spectrum of harmonic noise has the unique peak at the frequency
\begin{equation}
 \omega_{\mathrm{p}}=\sqrt{\omega_0^2-\frac{\Gamma^2}{2}}
\label{w_p}
\end{equation}
with the width
\begin{equation}
\Delta\omega = \sqrt{\omega_{\mathrm{p}}+\Gamma\omega'}-
\sqrt{\omega_{\mathrm{p}}-\Gamma\omega'},
 \label{width}
\end{equation}
where 
\begin{displaymath}
\omega'=\sqrt{\omega_0^2-\Gamma^2/4}.  
\end{displaymath}
One can easily find that 
\begin{displaymath}
 f(t)\to \sin(\omega_0t+\phi_0),
\end{displaymath}
as $\Gamma\to 0$. Setting $f(0)=1$, $y(0)=0$, and 
\begin{equation}
 \tau=\frac{\pi}{2\omega_0},
 \label{tau}
\end{equation}
one finds
\begin{equation}
V(x,t)=\sin(x-\omega_0t)  
\end{equation}
in the case of $\Gamma=0$. 
Hence, it turns out that $V(x,t)$
for $\Gamma>0$ behaves as a fluctuating plane wave \cite{PLA}. 

Owing to broken space-time symmetries, perturbation (\ref{Vxt}) can give rise to directed transport.
In the semiclassical regime and in the case of the periodic potential $U(x)$, 
direction of the current coincides with the direction of the perturbation phase velocity \cite{PLA}, that is,
the perturbation creates force that drags atoms towards $x\to\infty$. 
Transition of atoms from finite to infinite regime becomes possible due to noise-induced destruction
of dynamical barriers in classical phase space \cite{Order}.
More intricate behaviour is observed in the deep quantum regime, when interband tunneling is negligible,
and system dynamics is restricted to the lowest energy band.
As it was shown in \cite{Fromhold-PRA13}, current direction qualitatively depends on the perturbation amplitude.
For low amplitude values, current velocity grows with increasing of the amplitude, until it becomes equal to the phase 
velocity of the perturbation. However, as the amplitude of the perturbation exceeds some threshold value,
the current velocity rapidly decreases and changes its sign, that is, there appears transport in the opposite direction.
This phenomenon is closely related to the specific form of Bloch oscillations.
Addition of harmonic noise into the perturbation should, however, enhance interband transitions whereby violating
the single-band picture of motion.
This issue was addressed for the model of a driven tilted lattice \cite{Wimberger-NJP,Wimberger-FNL}.

Taking into account some similarity between perturbations used in our model and the model considered in \cite{Fromhold-PRA13}, it is reasonable to examine transport
properties for different values of the perturbation amplitude $\varepsilon$.
In the present work we consider two cases, $\varepsilon=0.05$ and $\varepsilon=0.25$, referring to them as weak and moderate driving, 
respectively.
Both these cases are considered in the next section by means of numerical simulation.

\section{Numerical simulation}
\label{Numer}

In the present section we study transport properties of cold atoms in the optical potential described in the preceding section.
We integrate numerically the Schr\"odinger equation (\ref{Shrod}) for the ensemble of 1000 realizations of the potential.
The initial condition is chosen in the Gaussian form
\begin{equation}
 \Psi(x,t=0)=C\exp\left[-\frac{(x-x_0)^2}{4\sigma_{x}^2(0)}\right],
\end{equation}
where $\sigma_{x}(0)=10\pi$, $x_0=0$, $C$ is the constant determined by the normalization condition
\begin{equation}
 \int |\Psi(x)|^2\,dx = 1.
 \label{norma}
\end{equation}
%

%1000 realizations of the potential $U(x) + \varepsilon V(x,t)$ were used for each parameter set.

%
\begin{figure}[h!tb]
\resizebox{0.45\textwidth}{!}{%
  \includegraphics{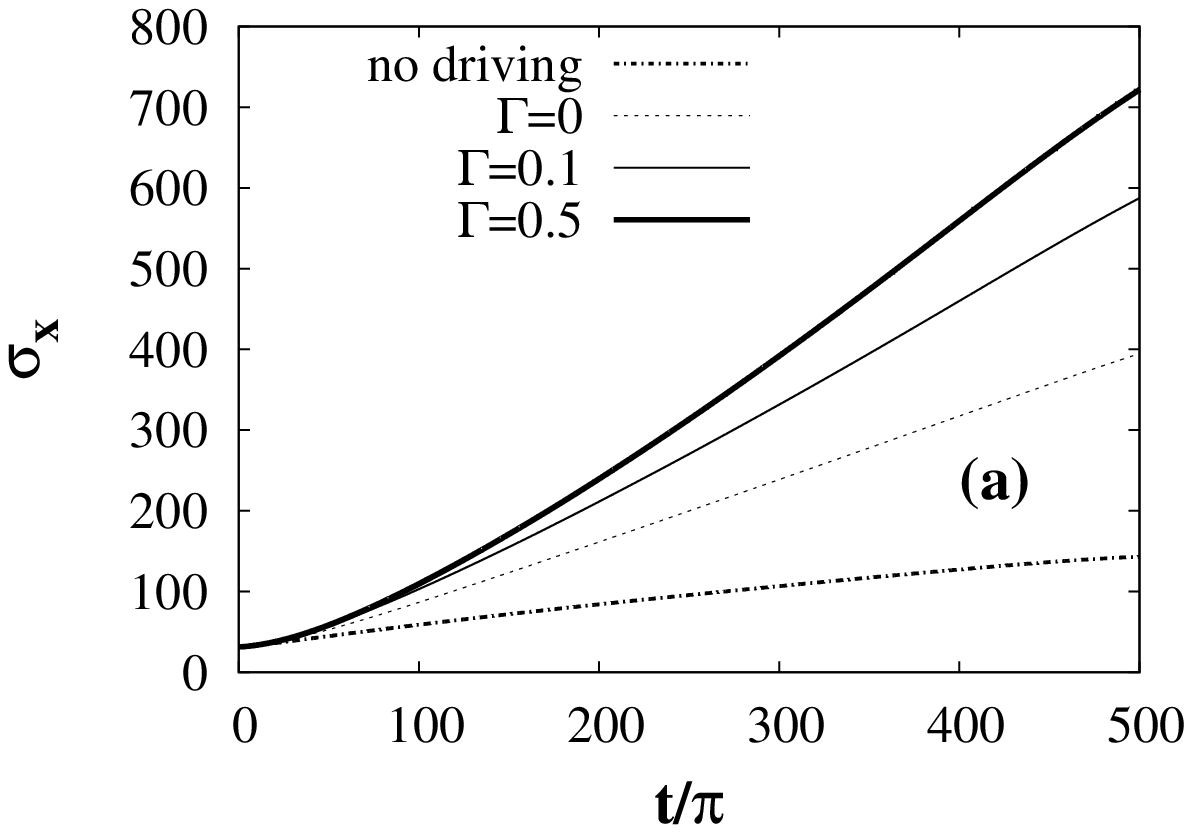}
}\\
\resizebox{0.45\textwidth}{!}{%
  \includegraphics{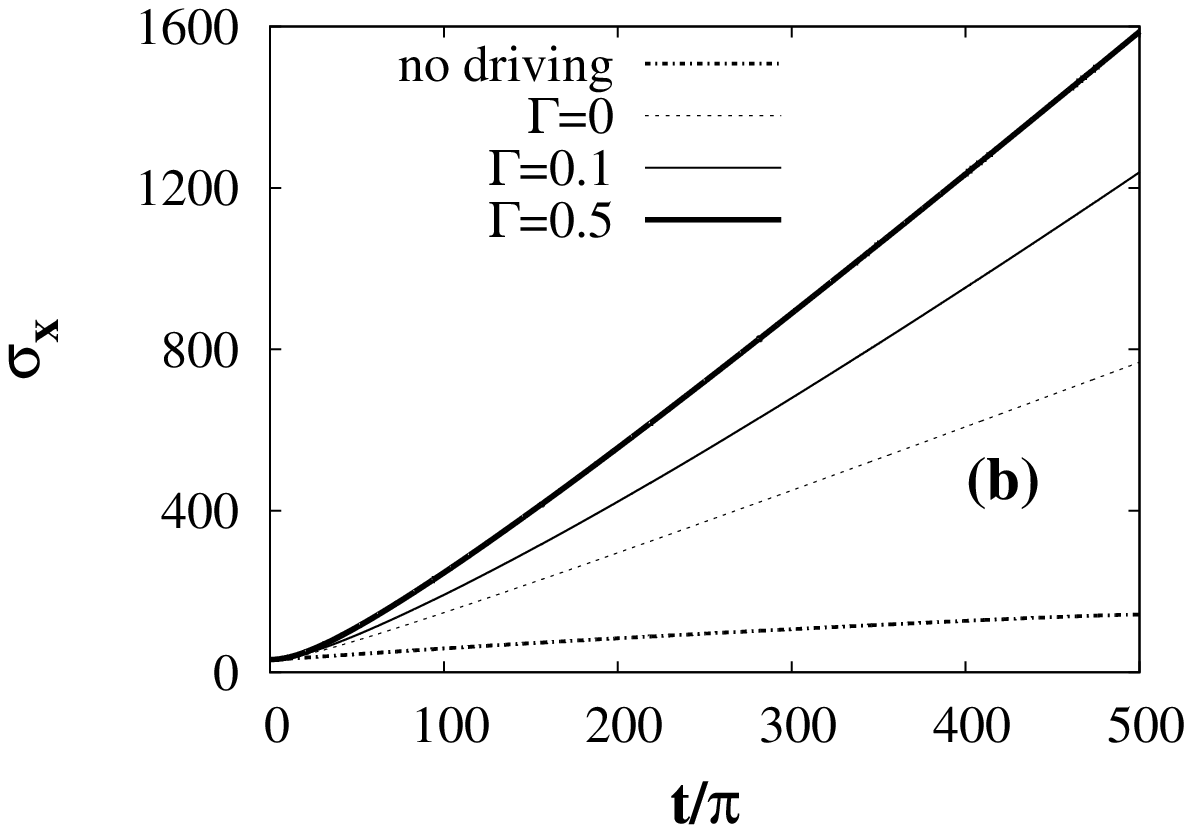}
}

\caption{
Ensemble-averaged position variance as function of time. 
(a) $\varepsilon=0.05$, (b) $\varepsilon=0.25$.
%Level spacing distributions in the energy bands corresponding  to 
%$E\le 1$ (dashes), $1<E\le 4$ (thick solid), and $4<E\le 10$ (thin solid).
} 
\label{fig-sgmx}
\end{figure}
\begin{figure}[h!tb]
\resizebox{0.45\textwidth}{!}{%
  \includegraphics{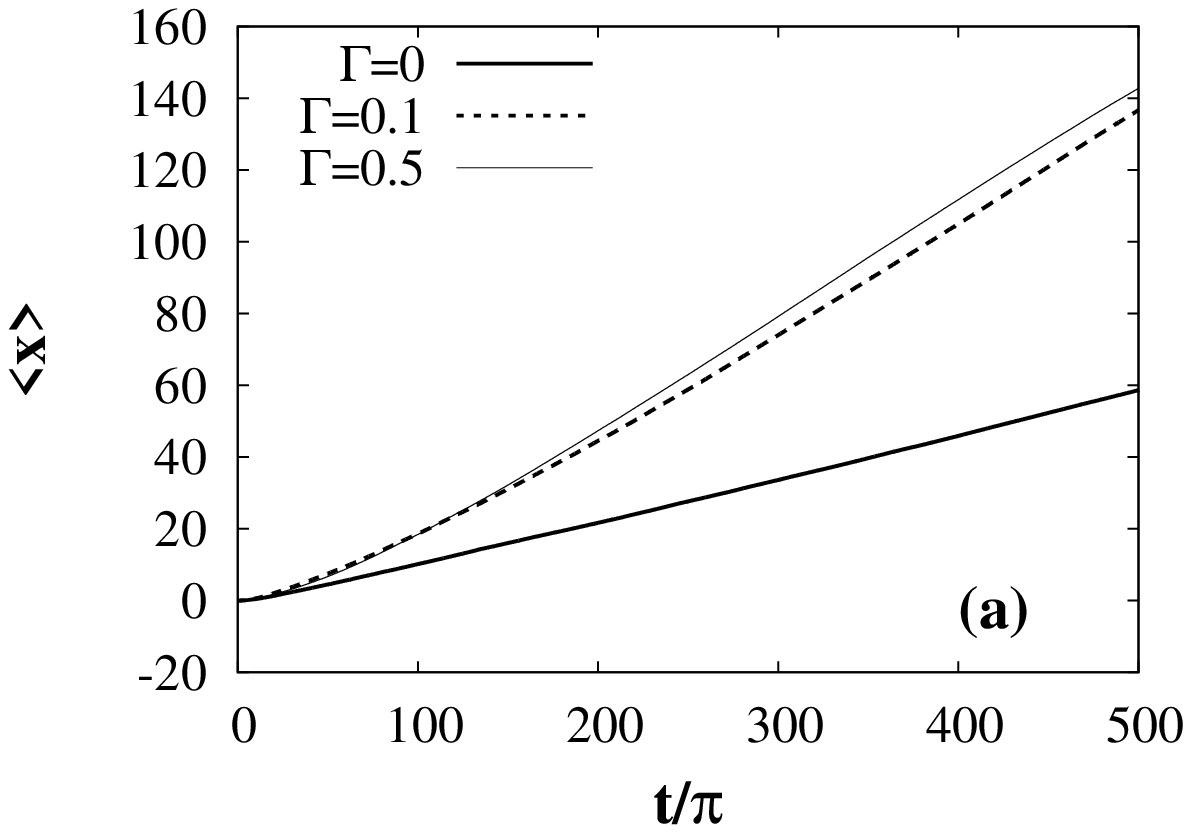}
}\\
\resizebox{0.45\textwidth}{!}{%
  \includegraphics{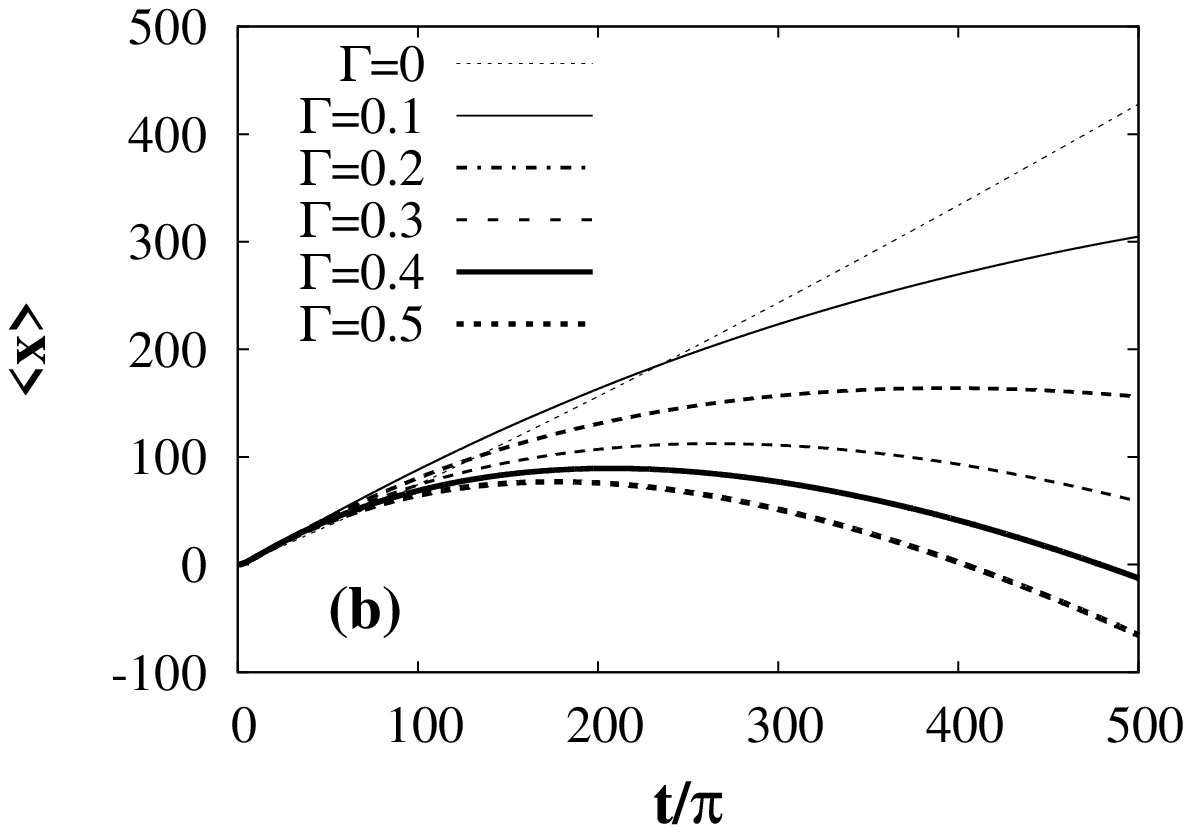}
}

\caption{Mean position as function of time. 
(a) $\varepsilon=0.05$, (b) $\varepsilon=0.25$.
} 
\label{fig-meanx}
\end{figure}

\subsection{Transport in the case of weak driving: dragging regime}
\label{Weak}

Let's begin with the discussion of numerical results corresponding to the case of weak driving $\varepsilon=0.05$.
Fig.~\ref{fig-sgmx}(a) demonstrates time dependence of ensemble-averaged position variance
\begin{equation}
\sigma_x=\frac{1}{\sqrt{J}}\sqrt{\sum\limits_{j=1}^J(Q_j-r_j^2)},
\label{sgm} 
\end{equation}
where $J=1000$ is the number of potential realizations,
$r_j$ is wavepacket displacement calculated 
with the $j$th realization of the potential
via the formula
\begin{equation}
 r_j = \int x|\Psi^{(j)}(x)|^2\,dx,
 \label{d}
\end{equation}
and $Q_j$ is squared displacement determined as
\begin{equation}
 Q_j = \int x^2|\Psi^{(j)}(x)|^2\,dx.
 \label{x2}
\end{equation}
%
%Three different values of the noise parameter $\Gamma$ are considered.
Hereafter $\Psi^{(j)}(x)$ means the solution of the Schr\"odinger equation (\ref{Shrod}) with $j$th realization of the potential.
The curve corresponding to the case $\varepsilon=0$ is also plotted for comparison.
Time-dependent perturbation significantly enhances wavepacket spreading as compared to the undriven case 
for all values of the noise parameter $\Gamma$.
Linear growth of $\sigma_x$ indicates excitation of ballistic states. 
The rate of spreading increases with increasing of $\Gamma$.
So, it turns out that broadening of the perturbation's spectral band enhances heating of atoms.
Fast spreading is accompanied by relatively slow drift of a wavepacket towards $x\to\infty$,
i.~e. in the direction of the phase velocity of the perturbation.
The drift is illustrated in Fig.~\ref{fig-meanx}(a) representing time dependence of mean position determined as
\begin{equation}
\aver{x} = \frac{1}{J}\sum\limits_j^J r_j.
 \label{mx}
\end{equation}
Thus, we observe dragging of atoms by the wave-like perturbation.
Notably, directed flux in the presence of noise is much larger than in the purely deterministic case $\Gamma=0$
because noise
leads to more extensive transitions between localized and ballistic states.
However, the rate of drift is nearly the same for $\Gamma=0.1$ and $\Gamma=0.5$, as well as for intermediate values
of $\Gamma$ (not shown). It means that enhancement of heating is partially supressed by fluctuations of the perturbation wave.

\begin{figure}[h!tb]
\resizebox{0.45\textwidth}{!}{%
  \includegraphics{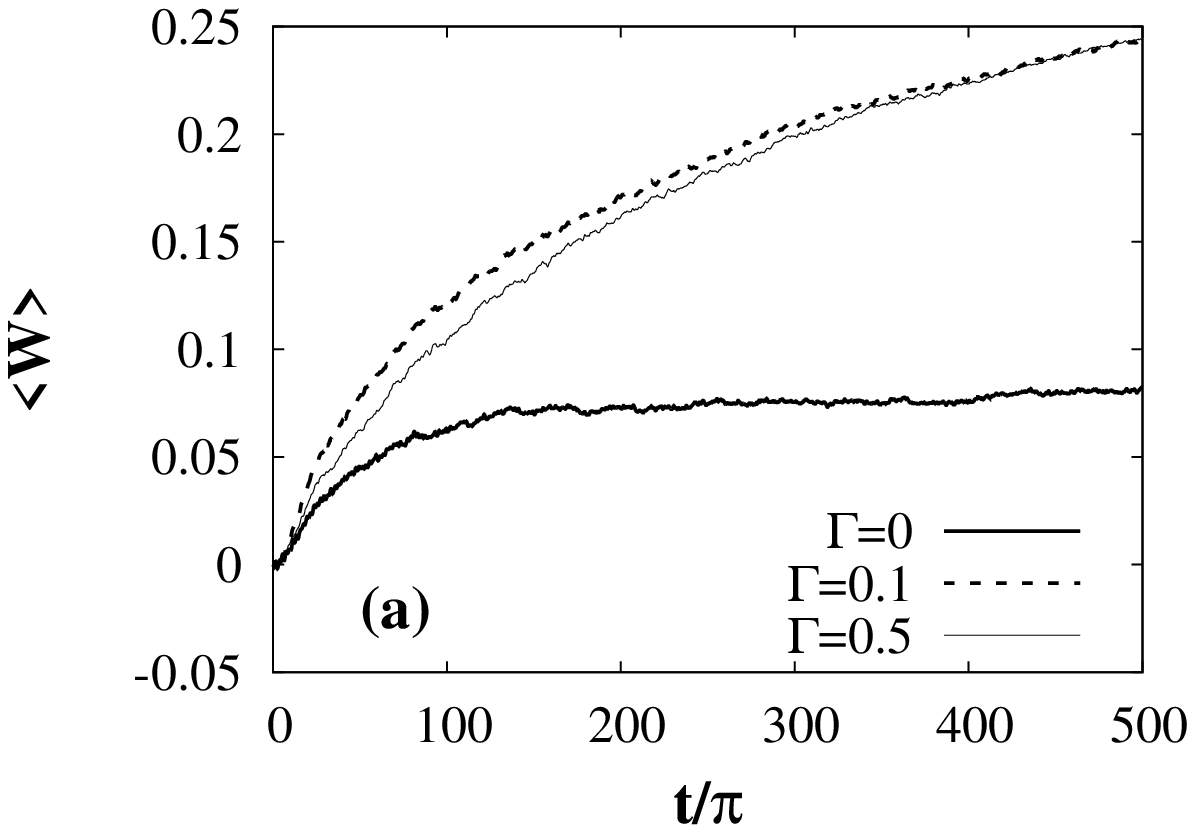}
}\\
\resizebox{0.45\textwidth}{!}{%
  \includegraphics{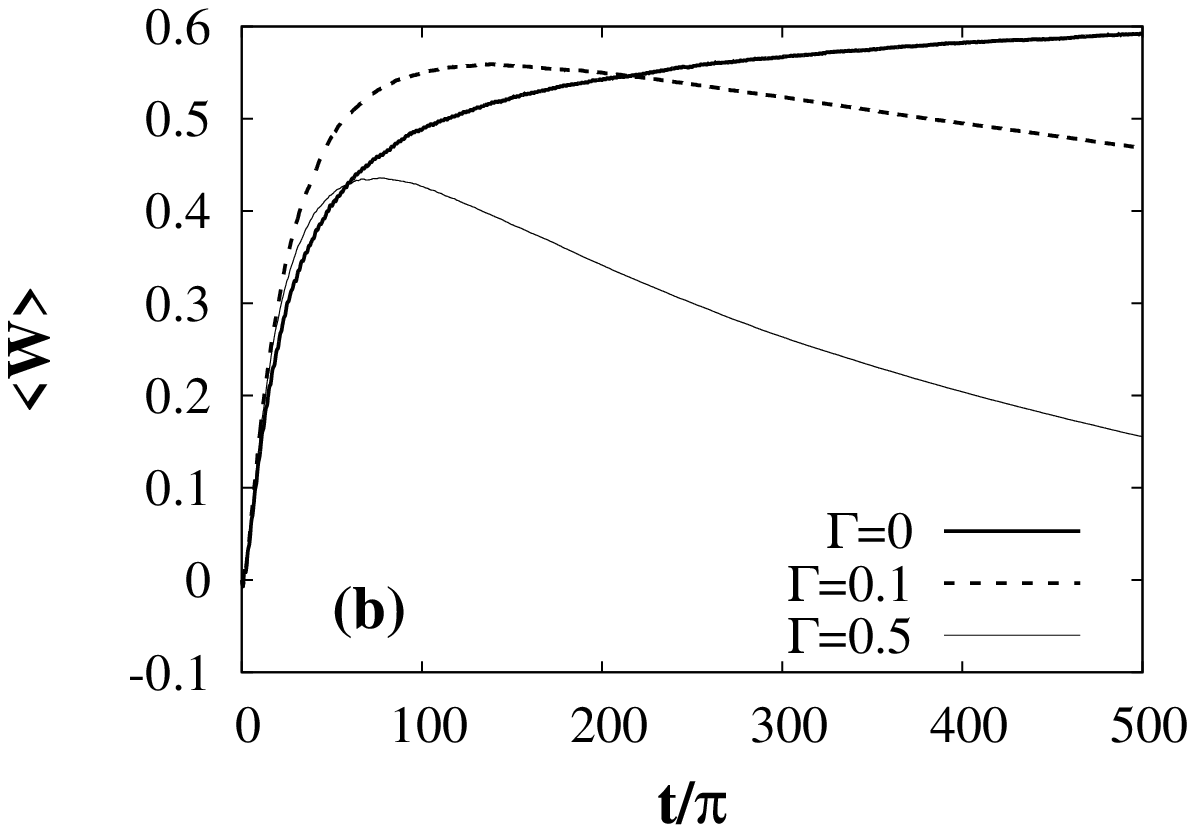}
}

\caption{Ensemble-averaged spatial population imbalance as function of time.
(a) $\varepsilon=0.05$, (b) $\varepsilon=0.25$.
%Level spacing distributions in the energy bands corresponding  to 
%$E\le 1$ (dashes), $1<E\le 4$ (thick solid), and $4<E\le 10$ (thin solid).
} 
\label{fig-W}
\end{figure}

It is also informative to consider time dependence
of the ensemble averaged spatial imbalance defined as
\begin{equation}
 \aver{W} = \frac{1}{J}\sum\limits_{j=1}^J (W_j^{\text{+}} - W_j^{\text{-}}),
 \end{equation}
where
 \begin{equation}
 \begin{aligned}
 W_j^{\text{+}} &= \int\limits_{0}^{\infty} |\Psi^{(j)}(x)|^2\,dx,\\
 W_j^{\text{-}} &= \int\limits_{-\infty}^{0} |\Psi^{(j)}(x)|^2\,dx.
\end{aligned}
 \end{equation}
Figure~\ref{fig-W}(a) shows that the strongest growth of spatial population imbalance is observed within
the initial time period, and then the growth becomes slower, 
although the imbalance remains far from the maximally accessible value 1.
The behaviour of $\aver{W}(t)$ reflects the process
of energy tranfer from localized states to ballistic ones having certain momentum value.
It turns out that states with both negative and positive momentum values are excited,
with relatively small prevailence of the latter ones.
As the fraction of localized states decreases with time, the excitation weakens, and growth of $\aver{W}$
becomes slower.

\begin{figure}[h!tb]
\resizebox{0.45\textwidth}{!}{%
  \includegraphics{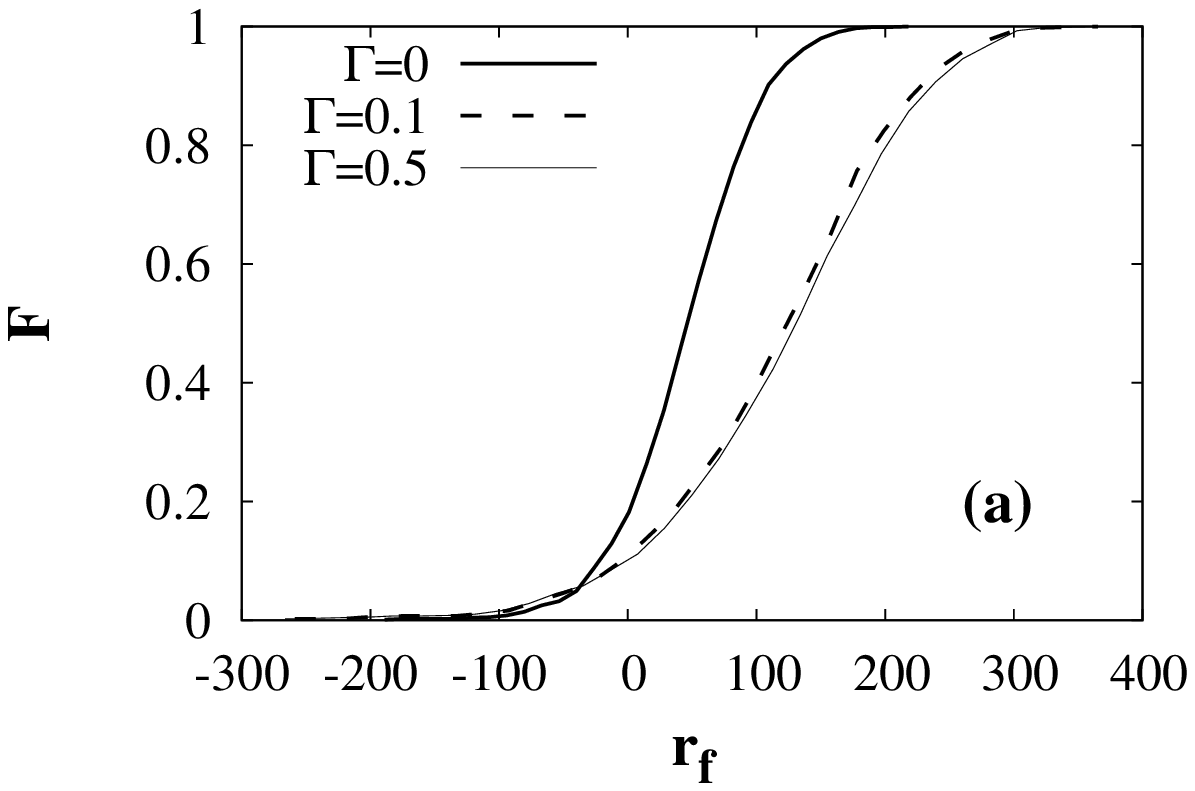}
}\\
\resizebox{0.45\textwidth}{!}{%
  \includegraphics{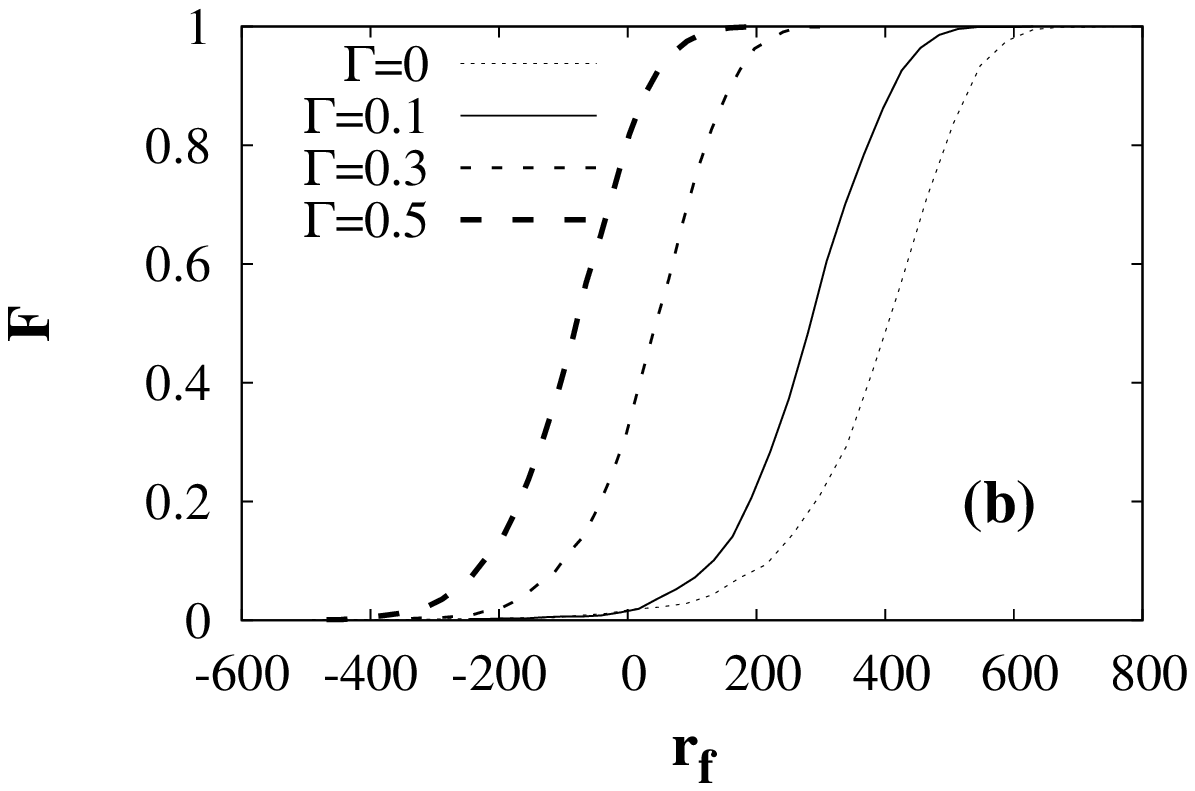}
}

\caption{
Cumulative distribution of wavepacket displacements at $t=500\pi$.
(a) $\varepsilon=0.05$, (b) $\varepsilon=0.25$.
%Level spacing distributions in the energy bands corresponding  to 
%$E\le 1$ (dashes), $1<E\le 4$ (thick solid), and $4<E\le 10$ (thin solid).
} 
\label{fig-fd}
\end{figure}

It is important to emphasize that excitation of ballistic states is a stochastic process, therefore,
transport properties for some single realization can deviate significantly from the picture
drawn by statistical averaging.
To illustrate it, we calculate the function
\begin{equation}
F(r_{\text{f}}) = \int\limits_{-\infty}^{r_{\text{f}}}\rho_{\text{r}}(r')\,dr',
 \label{Fd}
\end{equation}
being the cumulative distribution of displacement values at $t = 500\pi$
for an ensemble of potential realizations.
Fig.~\ref{fig-fd}(a) shows that nearly 20 percents of realizations exhibit prevailence 
of transport in the opposite direction to the dragging in the case of $\Gamma=0$.
Increasing of $\Gamma$ allows one to reduce fraction of such ``anomalous''
realizations to nearly 10 percents.
Thus, onset of directed current
with controllable direction is possible only with probability not equal to one.

\subsection{Transport in the case of moderate driving: onset of current reversal}
\label{Moder}

Now let's consider the case of the moderate driving, $\varepsilon=0.25$.
According to Fig.~\ref{fig-sgmx}(b), there is also ballistic spreading of a wavepacket,
but the rate of spreading is increased approximately two times as compared with the case of weak driving.
As in the case of weak driving, addition of noise remarkably enhances spreading. 

Time dependence of mean position looks in qualitatively different way.
In the noiseless case $\Gamma=0$, we observe the same regime of dragging as in the case of weak driving, with significantly increased
current velocity. However, as noise is added, the behaviour changes drastically. 
Dragging persists only until some time horizon, and then current changes direction.
It is clearly demonstrated in Fig.~\ref{fig-meanx}(b). The time of current reversal decreases with increasing of $\Gamma$.
Comparing Figs.~\ref{fig-meanx}(b) and \ref{fig-W}(b), one can deduce that, despite of the current reversal,
the majority of atomic states remain in the range of positive values of $x$, albeit their fraction decreases with time.
It means that backward current is produced by progressive accumulation of atomic states with large negative velocity.

Onset of current reversal is also reflected in the distributions of wavepacket displacements at $t=500\pi$, demonstrated in Fig.~\ref{fig-fd}(b).
In the noiseless case $\Gamma=0$, almost all realizations of potential give rise to positive displacements.
Inclusion of fluctuations increases probability of backward displacement. 
For $\Gamma=0.5$, backward displacements dominate the overall statistics.

Thus, numerical simulation exhibits a somewhat unusual phenomenon of current reversal that occurs if the noise parameter
$\Gamma$ is non-zero.
Theoretical explanation of this phenomenon is given in the next section.

\section{Dynamics in the momentum space}
\label{Momentum}

The origin of the current reversal observed can be found out if we
examine wavepacket evolution in the space of of momentum eigenstates
\begin{equation}
 |m>=\psi_m = \frac{1}{\sqrt{L}}e^{(i/\hbar)(p_mx-E_mt)},
\end{equation}
where
\begin{equation}
\begin{aligned}
 p_m = \frac{2\pi m\hbar}{L},\quad E_m = \frac{p_m^2}{2M},\\ m = -m_0,-m_0+1,...,m_0-1,m_0.
\end{aligned}
 \end{equation}
Temporal evolution of occupation probabilities 
\begin{equation}
 \rho_m=\left|\int \psi_m^*\Psi\,dx
 \right|^2
\end{equation}
can be described by the master equations \cite{Wilkinson_Mehlig_Cohen}
\begin{equation}
\frac{d\rho_l}{dt}=\sum\limits_m G_{lm}(\rho_m-\rho_l),
\label{master} 
\end{equation}
where $G_{lm} = |H_{lm}|^2$, where $H_{lm}$ is the matrix element responsible for transition between the momentum states $l$ and $m$.
$H_{lm}$ can be represented as a sum
\begin{equation}
 H_{lm} = H_{lm}^{a} + H_{lm}^{b} + H_{lm}^{c},
\end{equation}
where
\begin{equation}
 H_{lm}^{a} = \frac{f(t)e^{i\omega_{lm}t}}{L}\int\limits_0^L e^{i\Delta p_{lm}x/\hbar}\sin{x}\,dx,
\end{equation}
\begin{equation}
 H_{lm}^{b} = \frac{f(t+\tau)e^{i\omega_{lm}t}}{L}\int\limits_0^L e^{i\Delta p_{lm}x/\hbar}\cos{x}\,dx,
\end{equation}
\begin{equation}
 H_{lm}^{c} = \frac{e^{i\omega_{lm}t}}{L}\int\limits_0^L U(x)e^{i\Delta p_{lm}x/\hbar}\,dx,
\end{equation}
\begin{equation}
 \Delta p_{lm}=p_{l}-p_{m},\quad
 \omega_{lm}=(E_{l}-E_{m})/\hbar.
\end{equation}
After integration, we find
\begin{equation}
 H_{lm}^{a} = \Biggl\{\Biggr.
 \begin{aligned}
  &\frac{i\,\text{sgn}(\Delta p_{lm})}{2}f(t)e^{-i\omega_{lm}t},\quad &\Delta p_{lm}\pm 1,\\
   &0,\quad &\Delta p_{lm}\ne 1
\end{aligned}
\label{Ha0}
\end{equation}
\begin{equation}
 H_{lm}^{b} = \Biggl\{\Biggr.
 \begin{aligned}
   &\frac{1}{2}f(t+\tau)e^{-i\omega_{lm}t},
   \quad &\Delta p_{lm}\pm 1,\\
   &0,\quad &\Delta p_{lm}\ne 1
\end{aligned}
\label{Hb0}
\end{equation}
\begin{equation}
 H_{lm}^{c} = \exp(-i\omega_{lm}t)U_{lm},
\end{equation}
Terms $H_{lm}^{a}$ and $H_{lm}^{b}$ correspond to resonant transitions under action
of the time-dependent perturbation $V(x,t)$.
It is reasonable to average them over sufficiently long time interval in order to eliminate 
short-time interference effects. 
This procedure corresponds to calculation of the transition amplitudes by means of the selebrated Fermi's golden rule.
The averaged terms are expressed as
\begin{equation}
 H_{lm}^{a} = \frac{i}{2T}\text{sgn}(\Delta p_{lm})\int\limits_0^T f(t)\exp(-i\omega_{lm}t)\,dt,
 \label{Ha1}
\end{equation}
\begin{equation}
 H_{lm}^{b} = \frac{1}{2T}\int\limits_0^T f(t+\tau)\exp(-i\omega_{lm}t)\,dt.
 \label{Hb1}
\end{equation}
$f(t)$ and $f(t+\tau)$ can be represented as Fourier integrals
\begin{equation}
 f(t) = \frac{1}{2\pi}\int F(\omega)e^{i\omega t}\,d\omega,
 \label{ft}
\end{equation}
\begin{equation}
 f(t+\tau) = \frac{1}{2\pi}\int F'(\omega)e^{i\omega t}\,d\omega.
 \label{ftt}
\end{equation}
Substituting (\ref{ft}) and (\ref{ftt}) into (\ref{Ha1}) and (\ref{Hb1}), respectively, and taking
the limit $T\to\infty$, we find
\begin{equation}
 H_{lm}^{a} = \frac{i}{2}\text{sgn}(\Delta p_{lm})F(\omega=\omega_{lm}),
\end{equation}
\begin{equation}
 H_{lm}^{b} = \frac{1}{2}F'(\omega=\omega_{lm}/\hbar).
\end{equation}
Only the resonant contribution is taken into account in these equations.
Assuming that time of phase correlations of $f(t)$ is large compared to $\tau$,
we can use approximation
\begin{equation}
  F'(\omega=\omega_{lm}) \simeq e^{i\omega_{lm}\tau}F\left(\omega=\omega_{lm}\right).
\end{equation}
Validity of this approximation requires that noise bandwidth $\Delta\omega$ should not be large. It is satisfied for moderate values
of the noise parameter $\Gamma$.

Term $H_{lm}^{c}$ corresponds to transitions caused by incoherent scattering on the random potential $U(x)$.
These transitions lead to broadening of the momentum spectra and remarkably affect wavepacket spreading.
However, this term is not resonant, and, after some finite time interval, the transitions it causes weaken due to interference.
In addition, the scattering process doesn't influence the directivity of transport because left-going and right-going states 
are created with equal probabilities.
As the transport directivity is our major concern, we can 
take into account the effect of broadening by means of proper choice of initial conditions
in (\ref{master}), while the contribution of the term $H_{lm}^{c}$ into rate constants
can be eliminated by means of averaging over time.
 It implies that qualitative (but not quantitative) description of directed current variability can be obtained in terms of 
a reduced model that doesn't involve the disordered potential.
Thus, the time-averaged matrix element reads
\begin{equation}
%\begin{aligned}
\bar H_{lm}= F(\omega=\omega_{lm})
 \left[\frac{i}{2}\text{sgn}(\Delta p_{lm})+e^{i\omega_{lm}\tau}
 \right].
 \end{equation}
Hence, we can find the corresponding transition rate
\begin{equation}
\begin{aligned}
 G_{lm}= \varepsilon^2 S(\omega=\omega_{lm})
 \Biggl[\Biggr.&\cos^2{\omega_{lm}\tau}+\\
 &+\left(\sin{\omega_{lm}\tau} + \frac{\text{sgn}(\Delta p_{lm})}{2}\right)^2
 \Biggl.\Biggr],
\end{aligned} 
\label{Glm}
\end{equation}
where $S(\omega)$ is harmonic noise power spectrum given by \cite{HN}
\begin{equation}
 S(\omega)=\frac{\beta\Gamma}{\pi[\omega^2\Gamma^2 + (\omega^2-\omega_0^2)^2]}.
 \label{hn-sp}
\end{equation}
Formula (\ref{Glm}) infers that transitions in the halfspaces corresponding
to negative and positive momentum values have different rates.
As spectral width of harmonic noise is relatively small,
we can use approximation
\begin{equation}
|\omega_{lm}|\approx\omega_0. 
\end{equation}
Then, taking into account (\ref{tau}), 
one can replace $\omega_{lm}\tau$ as $\text{sgn}(\omega_{lm})\pi/2$ and simplify (\ref{Glm}) as
\begin{equation}
G_{lm}\approx \varepsilon^2 S(\omega=\omega_{lm})\left[
\text{sgn}(\omega_{lm})+\frac{\text{sgn}(\Delta p_{lm})}{2}
\right]^2
\label{Glm-s} 
\end{equation}
We have $\text{sgn}(\omega_{lm})=-\text{sgn}(\Delta p_{lm})$ for left-going states
and $\text{sgn}(\omega_{lm})=\text{sgn}(\Delta p_{lm})$ for right-going ones.
This means that transitions between left-going momentum states
are much less extensive.
Since only limited range of momentum values corresponds to strongly coupled states,
there should be accumulation of left-going states which have larger lifetimes.

\begin{figure}[h!tb]
\resizebox{0.45\textwidth}{!}{%
  \includegraphics{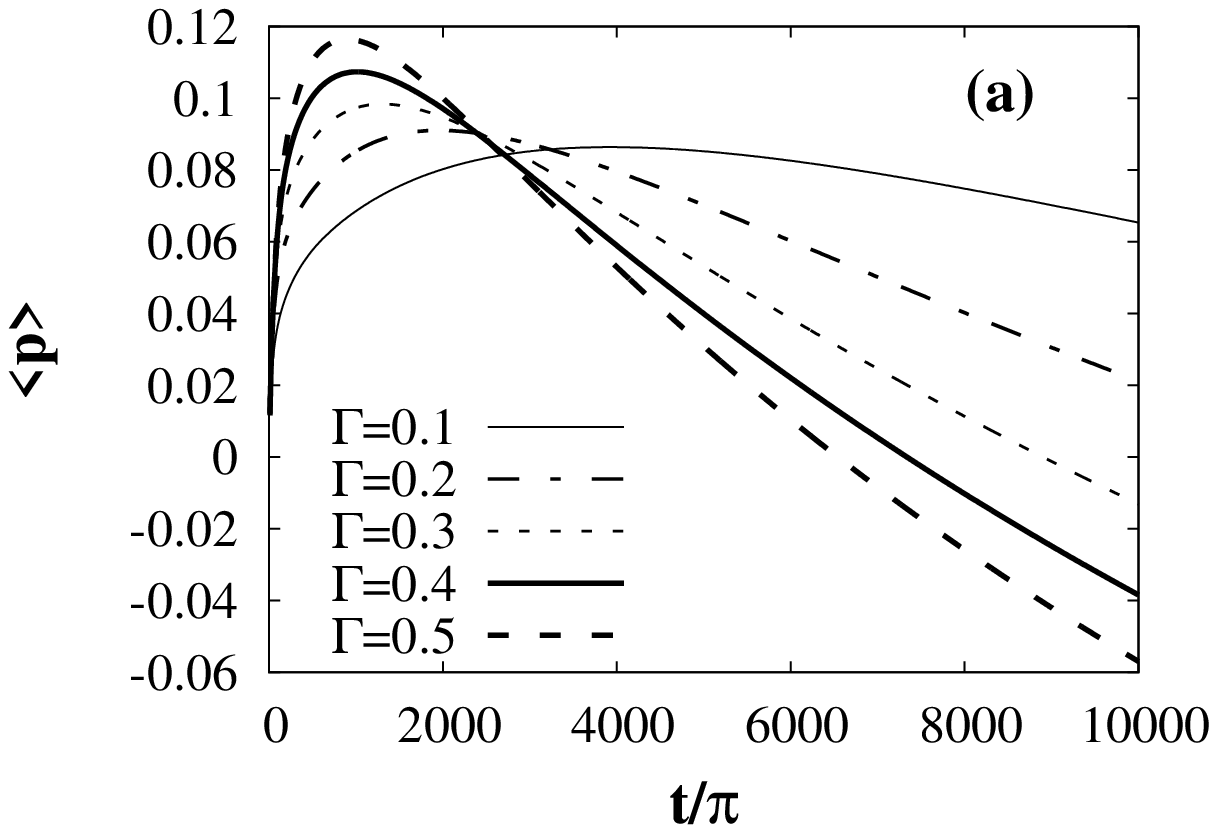}
}\\
\resizebox{0.45\textwidth}{!}{%
  \includegraphics{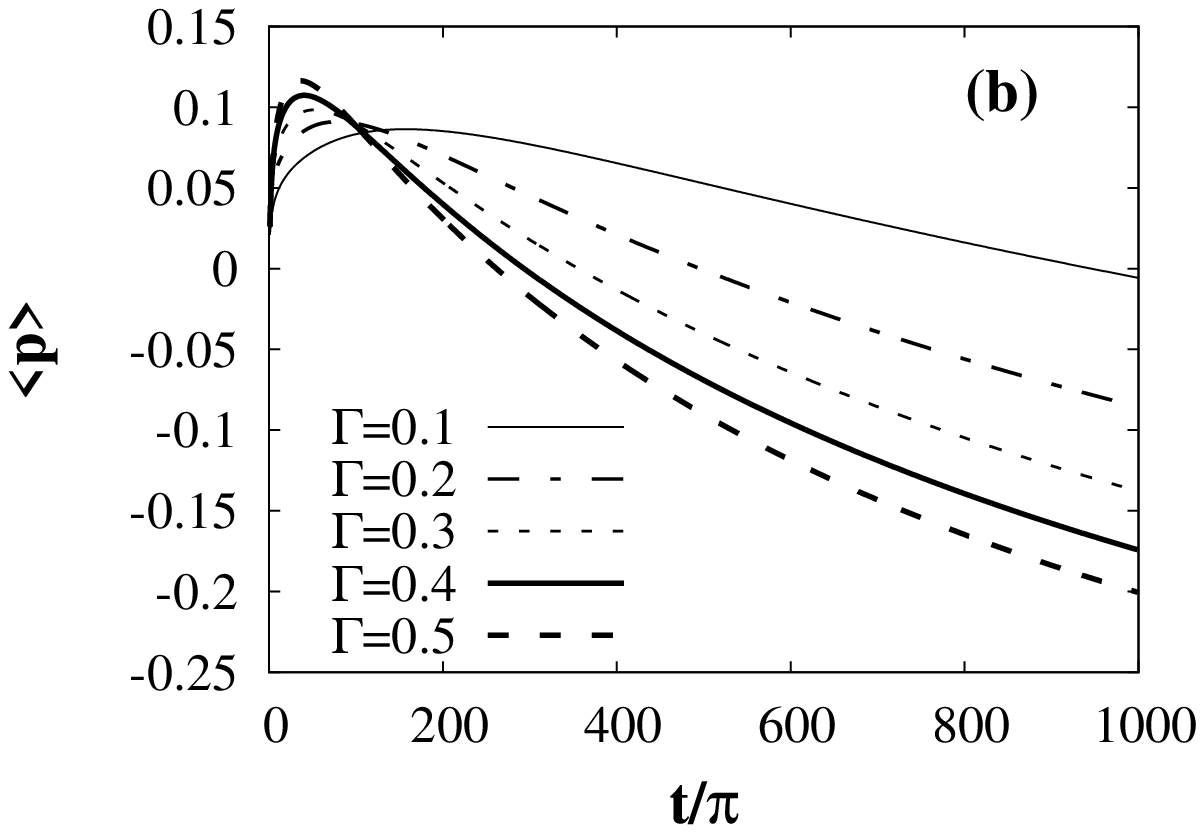}
}

\caption{Mean momentum calculated by solving equations (\ref{master}) as function of time. 
(a) $\varepsilon=0.05$, (b) $\varepsilon=0.25$.
%Level spacing distributions in the energy bands corresponding  to 
%$E\le 1$ (dashes), $1<E\le 4$ (thick solid), and $4<E\le 10$ (thin solid).
} 
\label{fig-meanp}
\end{figure}

To verify the above suggestion, we solved numerically the system of master equations (\ref{master})
with the Gaussian initial condition
\begin{equation}
\rho_m = A\exp\left(-\frac{p_m^2}{2\sigma_p^2}\right),
 \label{initp}
\end{equation}
where 
\begin{equation}
 A = \frac{1}{\sum\limits_m \rho_m}.
\end{equation}
We used relatively large momentum variance $\sigma_p=1$ 
in order to mimick the effect of scattering on the disorder potential $U(x)$.
Fig.~\ref{fig-meanp} represents dependence of mean momentum on time for various values
of $\varepsilon$ and $\Gamma$.
Comparison of Figs. \ref{fig-meanp}(a) and (b) reveals evident similarity between the cases
of $\varepsilon=0.05$ and $\varepsilon=0.25$.
Nevertheless, current reversal occurs in the former case on significantly longer timescales.  
Indeed,  after substitution of (\ref{Glm-s}) into (\ref{master})
one can eliminate $\varepsilon$ by rescaling time as
$t'=\varepsilon^2 t$.
Thus, it turns out that the plane-wave perturbation is able to drag particles
only within some limited time intervals before the reversal.
Reversals don't appear in data shown in Fig. \ref{fig-meanx}(a) because the time interval considered is too short.
Broadening of the perturbation spectrum increases number of efficiently coupled momentum states and results in more
extensive diffusion in the momentum space. 
Consequently, the reversal happens earlier.
This explains facilitation of transport with increasing $\Gamma$.
It should be noted that calculations for longer timescales reveal further reversals,
however, time interval between two succesive reversals rapidly grows.

\section{Summary}
\label{Summary}

The present work is devoted to a simple one-dimensional quantum model involving
a disordered potential and time-dependent perturbation in the form of a fluctuating plane wave.
This model can be realized experimentally with optically trapped cold atoms.
Also, it can serve as a toy model for studying phonon-induced charge transport in disordered wires.

The main result of the work is the onset of current reversals which occur for non-zero values
of the model parameter quantifying fluctuations of the time-dependent force.
It is shown that this effect is a consequence of diffusion inhomogeneity in the momentum space.
Enhancement of noise facilitates diffusion in the momentum space and
diminishes the time needed for the reversal onset.

It should be mentioned that influence of the plane-wave-like perturbation
 onto dynamics of quantum wavepacket was recently considered in \cite{Fromhold-PRA13}. In that paper, 
it was found that atomic current can change its sign as the perturbation amplitude increases.
Despite the results of \cite{Fromhold-PRA13} and this paper look similar,
the underlying mechanismes are different.
In the model considered in \cite{Fromhold-PRA13}, the plane-wave-like perturbation leads to excursion of quasimomentum inside the lowest 
energy band. 
In our model, dynamics is not restricted to the lowest band, moreover,
the crucial role  in the reversal onset is played  by 
inter-level transitions, that is, energy of atoms is not restricted by the first band.
It should be emphasized that the onset of current reversals in our model
is a noise-induced effect accompanied by energy growth and
heating of atoms.

It is important to note that onset of current reversals can be qualitatively described 
by means of the reduced kinetic model that doesn't take into account effect of the disordered
background potential, that is, disorder doesn't play crucial role for the reversals.
It means that such reversals can be readily observed in simpler models where potential
involves only fluctuating plane wave. 
We hope to address this issue in forthcoming works.
Another issue of interest is how the current reversal effect manifests itself under quantum-to-classical crossover.

%Stronger noise.

%Transition to semicalssical regime.

\section*{Acknowledgments}

This work is supported by grants from the Russian Foundation of Basic Research (project 12-02-31416), 
and the joint grant of the Far-Eastern and Siberian Branches of the Russian Academy of Sciences.

%\begin{figure*}
%\vspace*{5cm}       % Give the correct figure height in cm
%\end{figure*}
%

% BibTeX users please use
% \bibliographystyle{h-physrev3}
% \bibliography{biblio}

\end{document}